\documentclass[usenatbib]{mn2e}
\usepackage{psfig}
\usepackage{ifthen}

\def\ltsima{$\; \buildrel < \over \sim \;$}
\def\simlt{\lower.5ex\hbox{\ltsima}}
\def\gtsima{$\; \buildrel > \over \sim \;$}
\def\simgt{\lower.5ex\hbox{\gtsima}}
\def\kpc{{\rm\,kpc}}

\def\msun{{\rm\,M_\odot}}

\def\CompactFigs{1}

\def\s{\ifmmode \widetilde \else \~\fi}
\def\={\overline}

\def\spose#1{\hbox to 0pt{#1\hss}}

\def\lta{\mathrel{\spose{\lower 3pt\hbox{$\mathchar"218$}}
     \raise 2.0pt\hbox{$\mathchar"13C$}}}
\def\gta{\mathrel{\spose{\lower 3pt\hbox{$\mathchar"218$}}
     \raise 2.0pt\hbox{$\mathchar"13E$}}}
\def\Dt{\spose{\raise 1.5ex\hbox{\hskip3pt$\mathchar"201$}}}    
\def\dt{\spose{\raise 1.0ex\hbox{\hskip2pt$\mathchar"201$}}}    

\def\dotsfill{\leaders\hbox to 1em{\hss.\hss}\hfill}

\loadboldmathitalic 
\title[An exterior Galactic Ring]
{One Ring to Encompass them All:\\
A giant stellar structure that surrounds the Galaxy }
\author[Rodrigo Ibata, Michael Irwin, Geraint Lewis, Annette Ferguson, Nial Tanvir]
{R. A. Ibata$^{1}$, M. J. Irwin$^2$, G. F. Lewis$^3$, A. M. N. Ferguson$^4$,
N. Tanvir$^5$\\
$^{1}$
Observatoire de Strasbourg, 11, rue de l'Universit\'e, F-67000, Strasbourg, 
France\\
$^{2}$
Institute of Astronomy, Madingley Road, Cambridge, CB3 0HA, U.K.\\
$^{3}$
School of Physics, University of Sydney, NSW 2006, Australia\\
$^{4}$
Kapteyn Astronomical Institute, Postbus 800, 9700 AV Groningen, The
Netherlands\\
$^{5}$
Physical Sciences, Univ. of Hertfordshire, College Lane, Hatfield, AL10 9AB, UK}

\date{\today}
\begin{document} 
\maketitle 
\begin{abstract}
We present evidence  that the curious stellar population  found by the
Sloan Digital Sky Survey  in the Galactic anticentre direction extends
to other  distant fields that  skirt the plane  of the Milky  Way. New
data, taken with the INT  Wide Field Camera show a similar population,
narrowly aligned  along the line  of sight, but with  a Galactocentric
distance that changes from $\sim  15\kpc$ to $\sim 20\kpc$ (over $\sim
100^\circ$ on the sky).  Despite being narrowly concentrated along the
line of sight, the structure  is fairly extended vertically out of the
plane of the Disk, with  a vertical scale height of $0.75\pm0.04\kpc$.
This  finding suggests  that the  outer rim  of the  Galaxy ends  in a
low-surface brightness stellar ring.   Presently available data do not
allow us to ascertain the  origin of the structure. One possibility is
that it is  the wraith of a satellite galaxy  devoured long-ago by the
Milky  Way,  though  our  favoured  interpretation is  that  it  is  a
perturbation   of   the  disk,   possibly   the   result  of   ancient
warps. Assuming  that the Ring  is smooth and axisymmetric,  the total
stellar mass in  the structure may amount to  $\sim 2\times 10^8\msun$
up to $\sim 10^9\msun$.
\end{abstract}


\section{Introduction}

Due to our  location within the disk of the Milky  Way, studies of the
global structure of this galactic component are hampered by projection
problems, crowding, dust, and  the presence of intervening populations
(such as the Bulge). Nowhere is this so problematic as in the study of
the very  outer edge of the  disk. The advent of  the recent wide-area
infra-red  surveys  (e.g.   2MASS   and  DENIS)  have  alleviated  the
extinction problem,  but the other problems remain,  with the distance
ambiguity  being particularly limiting.   Even the  future astrometric
mission GAIA \citep{perryman} is unlikely to give us a full picture of
the Galactic disk, due to  telemetry limits in regions of high stellar
density.

Yet  the outer  regions of  galactic  disks are  important regions  to
study, as  they provide  important clues to  our understanding  of the
global    structure   and    formation    of   galaxies    \citep[see,
e.g.,][]{kruit}.   These  are the  least  self-gravitating regions  of
galactic disks, and the presence of  the dark matter halo can begin to
be felt at these radii.  The  flaring of the outer disk constrains the
dark   matter  fraction  in   these  regions   \cite[][and  references
therein]{olling}.   Perhaps the  most interesting  aspect of  the very
outermost edge  of the disk  is that it  is expected to be  young.  In
galaxy formation simulations  that contain a gas component  as well as
Cold Dark Matter, galaxy disks tend  to grow from the inside out, with
the most recently accreted gas settling  down onto the end of the disk
\citep{navarro}.   Ensuing  star-formation  in regions  of  sufficient
density produces young stars, leading to a primarily young, metal-poor
stellar population in these galactic extremities, though radial mixing
in the disk may smear this information out \citep{sellwood}.  However,
recent  simulations \citep{sommer-larsen}  show that  some  disks form
outside-in as  well as inside-out,  in agreement with  tantalising new
evidence that  indicates that the  outer disk of the  Andromeda galaxy
may  well be  old  \citep{ferguson01}.  Determining  the  age of  disk
populations at large  radius will provide a good  test of current disk
formation models.

An  interesting  recent  development  in  the  study  of  the  stellar
populations of  the outer disk has been  presented by \citet{newberg},
based on  Sloan Digital  Sky Survey photometry  of fields  towards the
Galactic anticentre direction.  \citet{newberg} find an overdensity of
F-colored  stars close  to  the Galactic  plane  in the  constellation
``Monoceros'', with  a narrow colour-magnitude sequence  that belies a
stellar population $\sim  11\kpc$ from the Sun and  $\sim 18\kpc$ from
the Galactic  centre. The narrow  magnitude spread implies  a distance
spread of about  $\sim 2\kpc$, despite the fact  that the structure is
seen over a  wide range above and below  the Galactic plane stretching
from $b  \sim -25^\circ$ to $b  \sim 20^\circ$ (i.e.  $-5.5\kpc  < z <
4.5\kpc$).

The analysis of \citet{newberg} suggested that this stellar population
was  a  very nearby  orbiting  Galactic  satellite.   Here we  present
evidence of  similar color-magnitude features in fields  taken as part
of a  survey of the Andromeda  galaxy with the  Isaac Newton Telescope
(INT), and as part of a public survey observed with the same telescope
entitled the INT Wide Field Survey (WFS).

\section{The Isaac Newton Telescope Wide Field Camera Surveys}

The INT WFS is an initiative by the UK and Dutch communities to devote
a  large fraction  of the  INT to  deep and  wide-field  surveys. Many
fields  have now  been observed  since 1998.   However,  the resulting
coverage at  the present  time is patchy,  with most time  having been
spent  in  large  extragalactic  surveys towards  the  Galactic  polar
caps. Table~1 is  a listing of suitable WFS  or WFC observations below
$|b| \simlt  50^\circ$. In  Figure~1 we display  an example of  one of
these  fields,  the  Elais   field  N1,  located  at  $\ell=85^\circ$,
$b=+44^\circ$,  which  shows the  normal  Galactic stellar  population
sequences.    In  contrast,  \citet{newberg}   have  shown   that  the
anticentre  (``Monoceros'')  region  shows  an additional  feature  in
colour-magnitude space (their Figure~12), with shape similar to a main
sequence that has a turnoff at $g'-r'=0.25$, $g' \sim 19.5$ (in the AB
system).  

\begin{figure}
\ifthenelse{\CompactFigs=0}
{\centerline{\psfig{figure=Gal_Ring.fig01.ps,angle=270,width=10cm}}}
{\centerline{\psfig{figure=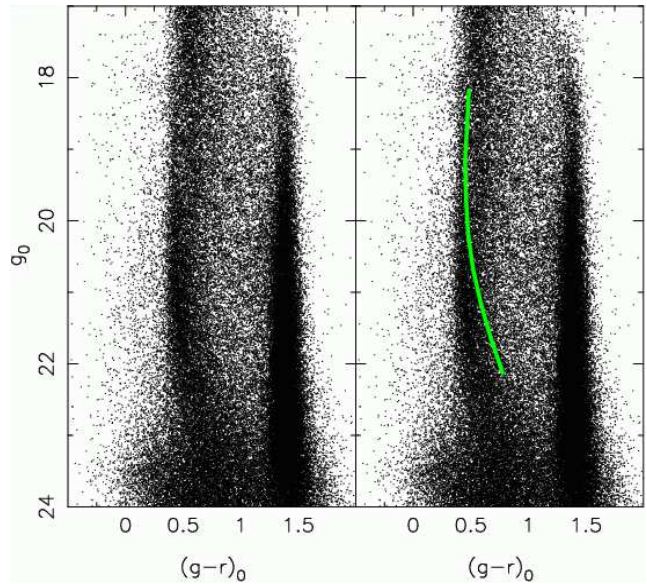,angle=0,width=\hsize}}}{}
\caption[]{The  colour-magnitude   diagram  of  the   Elais  field  N1
($\ell=85^\circ$,  $b=+44^\circ$),  which we  will  use  as a  control
field.  This comparison  region shows  the usual  Galactic components.
The Galactic disk dwarfs contribute to the well-populated red vertical
structure  at  ${\rm  (g-r)_0  \sim  1.4}$,  whereas  the  progressive
main-sequence turnoffs  of the  thick disk and  halo give rise  to the
blue vertical  structure at ${\rm (g-r)_0 \sim  0.5}$.  Eventually, at
magnitudes fainter than  ${\rm g_0 \sim 22}$ the  halo sequence curves
round  to  the  red  due  to  the  rapidly-falling  density  at  large
Galactocentric  distance.   (The  photometry  has been  corrected  for
extinction using the maps of \citealt{schlegel}). The right-hand panel
shows the same data as the left-hand panel, but we have superposed the
ridge-line of the structure of  interest in Figures~2--5 to serve as a
visual  aid in  the interpretation  of those  figures.  Note  that the
narrow stellar sequence  detected in fields Mono-N1 (Figures  2 and 3)
and in WFS-0801 (Figures 4 and 5), which lies close to the ridge line,
is not present in this comparison field.}
\end{figure}

In examining INT  WFC survey fields, we have  detected the presence of
this unexpected feature in other distant fields. Figure~2 displays the
colour-magnitude diagram  of the INT WFC field  ``Mono-N'' (located at
$\ell=150^\circ$,  $b=+20^\circ$); a population  that follows  a track
similar to  a narrow-main  sequence is seen  in addition to  the usual
Galactic  components.  This  sequence  is shown  more  clearly in  the
right-hand panel of Figure~3, in which we have used the Elais-N1 field
as a  ``background'' to subtract  off the normal  Galactic components.
Due  to the  difference in  Galactic latitude  between the  target and
control fields,  the thick disk  is not subtracted cleanly:  this poor
subtraction  of  the  thick  disk  is  seen as  a  smear  to  brighter
magnitudes and redder colours than the narrow sequence that delineates
the  abrupt faint end  of the  right hand  panel of  Figure~3. Another
INT WFS field that displays this excess  population is the field named
WFS-0801   ($\ell=180^\circ$,  $b=+30^\circ$),   the  colour-magnitude
diagram of  which is displayed in Figure~4.   Subtracting a background
estimated from the Elias-N1 field  gives the Hess diagram displayed on
the  right-hand panel  of  Figure~5. As  the  ``background'' field  is
closer in  Galactic latitude to  the WFS-0801 field,  this statistical
subtraction is much cleaner, allowing us to show the unexpected excess
population relatively free of contamination from the expected Galactic
components.

\begin{figure}
\ifthenelse{\CompactFigs=0}
{\centerline{\psfig{figure=Gal_Ring.fig02.ps,angle=270,width=10cm}}}
{\centerline{\psfig{figure=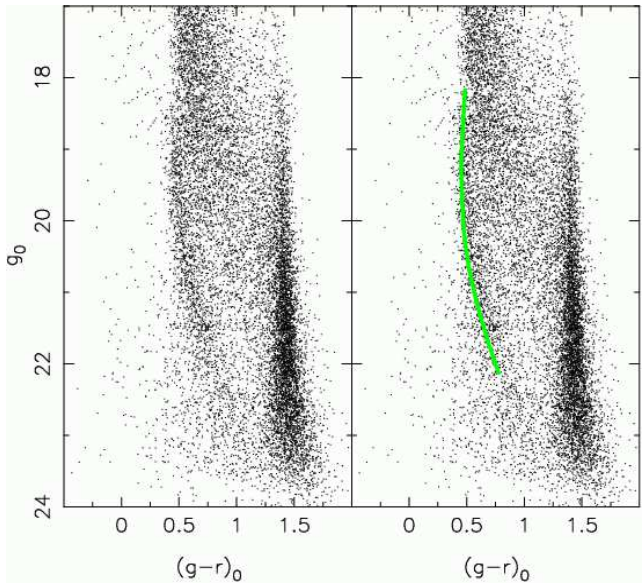,angle=0,width=\hsize}}}{}
\caption[]{The  colour-magnitude  diagram   of  a  field  (Mono-N)  at
$\ell=150^\circ$,   $b=+20^\circ$.   An   additional  colour-magnitude
feature is  present here over the  expected disk, thick  disk and halo
components, and is  seen as a narrow CMD structure,  similar to a main
sequence with  turn-off at  ${\rm (g-r)_0 \sim  0.5}$, ${\rm  g_0 \sim
19.5}$ (in the Vega system). Correcting for the difference between the
AB  and  Vega  photometry,  we  see that  the  peculiar  main-sequence
detected by  \citet{newberg} in Sloan  Digital Sky Survey  (SDSS) data
towards the Galactic anticentre is also clearly present in this field.
We used the colour transformations outlined in the text to convert the
ridge-line  of the feature  in the  SDSS S223+20  field; an  offset of
$-0.4$~magnitudes was needed to  match up the sequences, implying that
the structure  in the S223+20  field is more distant.   The right-hand
panel   shows  this  ridge-line   overlaid  on   the  colour-magnitude
diagram. The  similarity in  the turn-off colour  of this  feature and
that  of the  Galactic  thick disk  and  halo shows  that its  stellar
population is of comparable age to those ancient Galactic components.}
\end{figure}

\begin{figure}
\centerline{\psfig{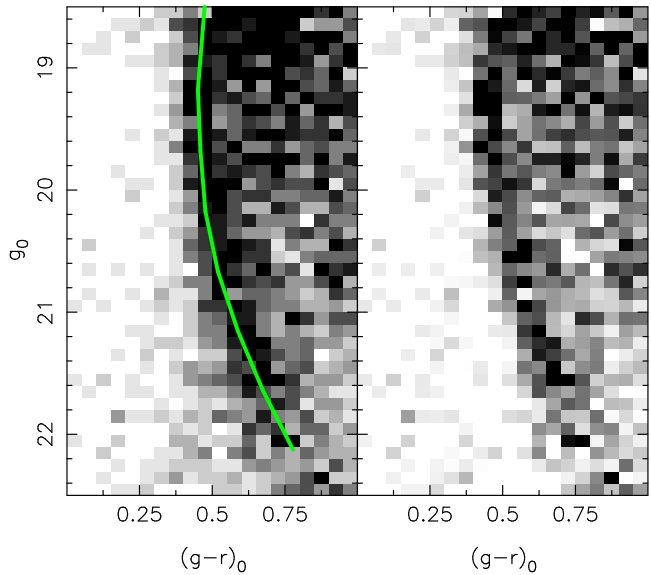}}
\caption[]{The left hand panel shows a (zoomed-in) Hess diagram of the
Mono-N  field  previously presented  in  Figure~2  (the ridge-line  of
Figure~2  has been  reproduced here  as well).   The  right-hand panel
shows  the result  of  subtracting the  Hess-diagram  of the  Elais-N1
comparison region  from the these data.  The  halo contribution, which
lies primarily at  fainter magnitudes than the ridge  line, is similar
in the two  fields, so the subtracted Hess  diagram is relatively well
cleaned of halo contaminants.  However, the comparison field (which is
located at $b=+44^\circ$) has a substantially lower surface density of
thick disk stars than the Mono-N field, so the thick disk contribution
is   only   slightly  reduced   in   the   subtracted  Hess   diagram.
Nevertheless,  the  narrow colour-magnitude  sequence  can be  clearly
perceived. The colour distribution  around the sequence shows a narrow
peak with $S/N>20$ (see Figure~9).}
\end{figure}

\begin{figure}
\ifthenelse{\CompactFigs=0}
{\centerline{\psfig{figure=Gal_Ring.fig04.ps,angle=270,width=10cm}}}
{\centerline{\psfig{figure=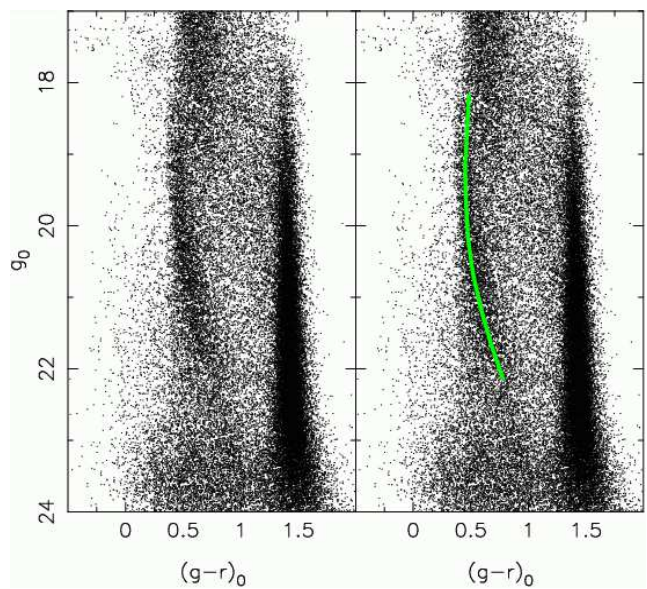,angle=0,width=\hsize}}}{}
\caption[]{As   Figure~2,  but   for   the  INT   WFS-0801  field   at
$\ell=180^\circ$,  $b=+30^\circ$. An  offset  of $-0.4$~magnitudes  is
needed to match up this sequence to that of the S223+20 field.}
\end{figure}

\begin{figure}
\centerline{\psfig{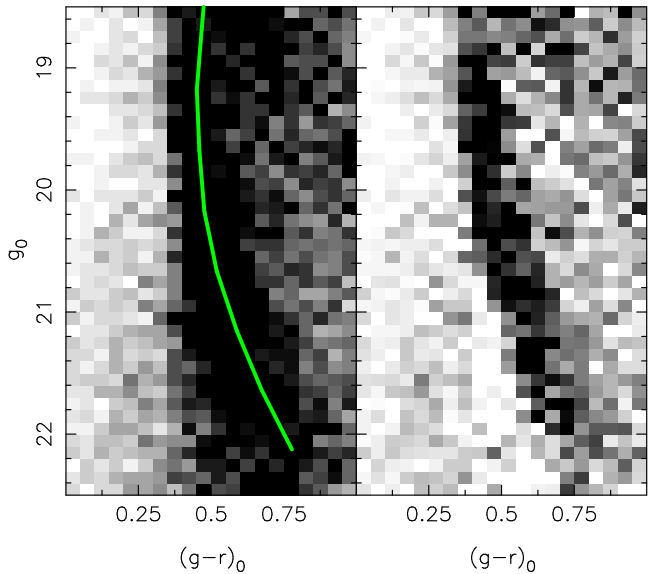}}
\caption[]{In a similar manner to Figure~3, the right-hand panel shows
the  Hess  diagram of  the  INT  WFS-0801  field at  $\ell=180^\circ$,
$b=+30^\circ$. The left-hand panel  displays the result of subtracting
the Elais-N1 comparison region from  the data in the right-hand panel.
Due  to  the  smaller  difference  in Galactic  latitude  between  the
WFS-0801 field  and the Elais-N1  field, the subtraction of  the thick
disk component is cleaner than  in Figure~3, and the excess population
stands out very clearly. Figure~9 shows that the excess is detected at
$S/N>30$.}
\end{figure}

The  other two proprietary  large surveys  \citep{ibata01, ferguson02,
ferguson03} were  conducted by our  group to reveal the  structure and
stellar   populations   in   the   halo  of   the   Andromeda   galaxy
($\ell=122^\circ$,    $b=-21^\circ$)   and    M33   ($\ell=134^\circ$,
$b=-31^\circ$).   A  further  two  fields  were observed  for  use  as
comparison  regions  for  these  surveys.  Table~1  also  lists  these
fields.   In  Figures~6  and  8  we display  the  ${\rm  V_0}$,  ${\rm
(V-i)_0}$, colour-magnitude  diagram in our M31 survey,  where we have
plotted separately the northern (lower $|b|$) and southern quadrants of
the survey.  The CMD structure seen in the SDSS ``Monoceros'' field is
also observed in our 30 square-degree field around M31.  The left-hand
panel of  Figure~7 shows  the Hess diagram  of the northern  M31 field
(note  that  this  is  a  zoomed-in view  of  Figure~6).   Lacking  an
appropriate background  field in $V$  and $i$ pass-bands, we  have been
forced  to use the  southern M31  field as  a comparison  region.  The
result of  subtracting the  southern M31 field  from the  northern M31
field is  displayed in  the right-hand panel  of Figure~7;  the excess
population is again clearly seen as a narrow sequence.

\begin{figure}
\ifthenelse{\CompactFigs=0}
{\centerline{\psfig{figure=Gal_Ring.fig06.ps,angle=270,width=10cm}}}
{\centerline{\psfig{figure=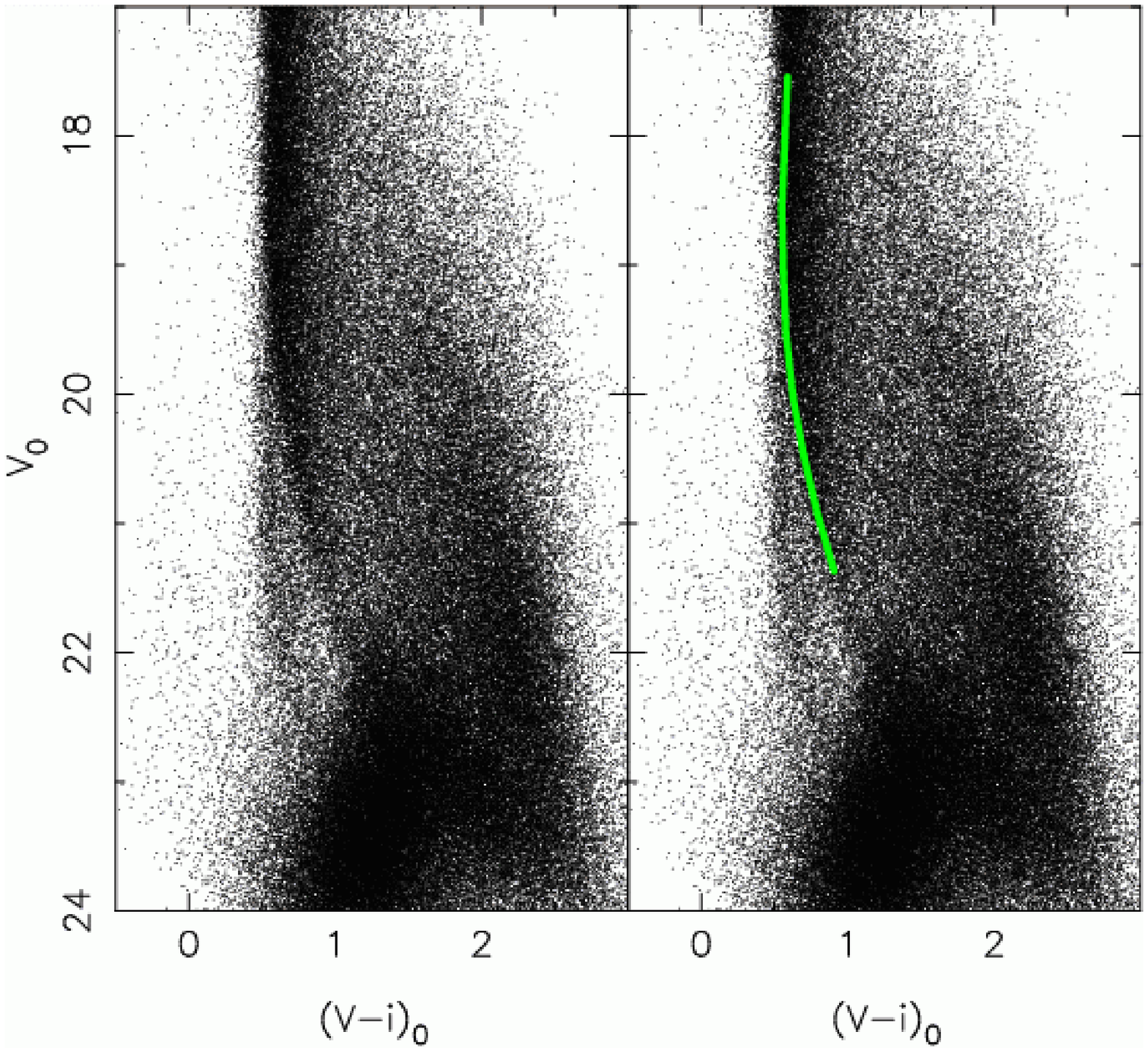,angle=0,width=\hsize}}}{}
\caption[]{The  colour-magnitude  diagram of  the  northern M31  field
($\ell=123^\circ$,  $b=-19^\circ$).  A  numerous  main-sequence shaped
colour-magnitude  structure is  seen  from ${\rm  (V-i)_0 \sim  0.5}$,
${\rm V_0 \sim  19}$ curving red-wards to ${\rm  (V-i)_0 \sim 1.0}$ at
${\rm  V_0 \sim  21.5}$.   Although this  diagram  cannot be  compared
directly to  Figures~2 or 4, or  to the SDSS CMD  of the ``Monoceros''
population, due to the  different photometric passbands, its behaviour
is  strikingly similar.  The  additional stars  with ${\rm  V_0 \simgt
22}$ are the top two magnitudes of the red giant branch in the halo of
M31.  The ridge-line of the  SDSS S223+20 feature, converted to (V,i),
requires a  significant offset of  $\sim -0.8$~magnitudes to  make the
two sequences coincide.  The right-hand panel shows this ridge-line.}
\end{figure}

\begin{figure}
\centerline{\psfig{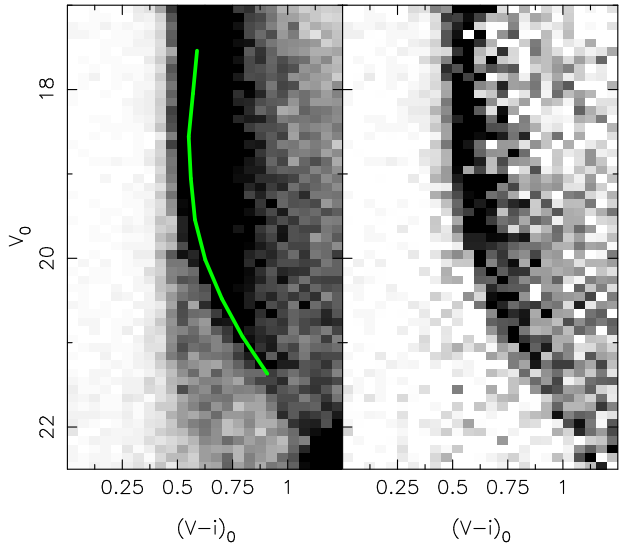}}
\caption[]{The left-hand panel shows  the Hess diagram of the northern
M31 field ($\ell=123^\circ$, $b=-19^\circ$).  This is a zoomed-in view
of Figure~6;  the ridge-line  of Figure~6 is  also reproduced  here to
guide the eye.  The right-hand  panel is the result of subtracting the
southern  M31  field   ($\ell=122^\circ$,  $b=-24^\circ$)  from  these
data. Given  that these two fields are  so close on the  sky, the halo
and thick  disk populations are  similar, so the contamination  of the
halo and  thick disk in the  Hess diagram on the  right-hand panel has
been   substantially   reduced   by   the  subtraction.   The   narrow
colour-magnitude  sequence stands  out with  $S/N>30$  (see Figure~9).
Note,  however,   that  the  southern   M31  field  is  not   a  good
``background'' region, since the excess population is also detected in
that field  (as seen in  Figures 8 and  10), though in  lower numbers.
This  means  that the  excess  population  is  over-subtracted on  the
right-hand panel, leading to an underestimate of its significance.}
\end{figure}

\begin{figure}
\ifthenelse{\CompactFigs=0}
{\centerline{\psfig{figure=Gal_Ring.fig08.ps,angle=270,width=10cm}}}
{\centerline{\psfig{figure=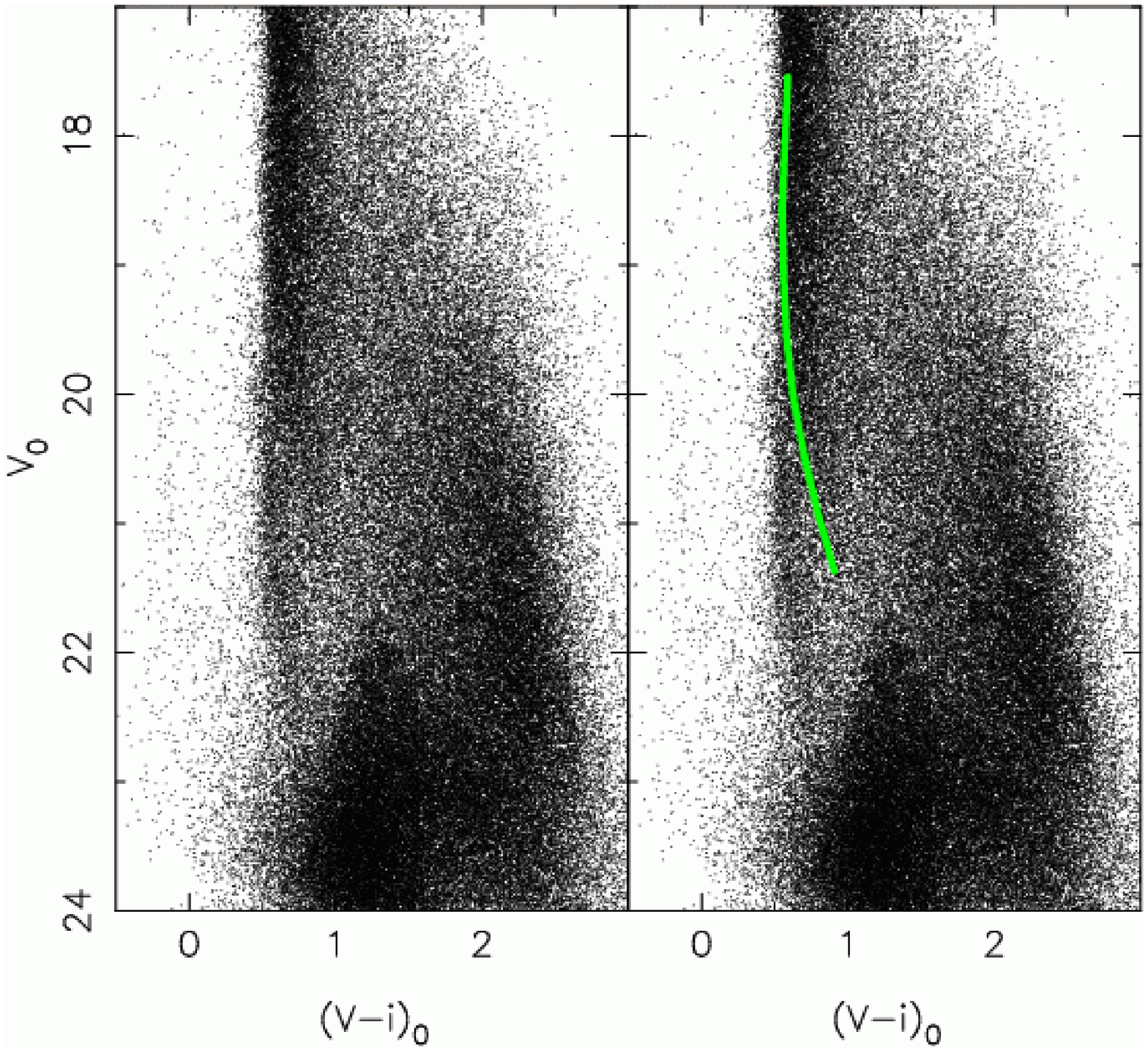,angle=0,width=\hsize}}}{}
\caption[]{As  Figure~6, but  for  the southern  M31  field that  lies
further from the Galactic plane ($\ell=122^\circ$, $b=-24^\circ$). The
``Monoceros''-like  population  is  still  present, though  much  less
numerous  than   in  the  northern   field  (which  is   displayed  in
Figure~6).  To  guide the  eye,  the  location  of the  ridge-line  in
Figure~6 is shown on the right-hand panel.}
\end{figure}

The  high statistical  significance  of the  detection  of the  excess
population in the different  fields is demonstrated in Figure~9, where
we show the distribution of  stars as a function of their displacement
in  colour  from the  ridge-lines.   Due to  the  lack  of a  suitable
background field, we  are unable to provide a  similar estimate of the
significance of  the detection in  the southern M31 field,  though the
population is clearly present in that field, as we show in Figure~10.

\begin{figure}
\centerline{\psfig{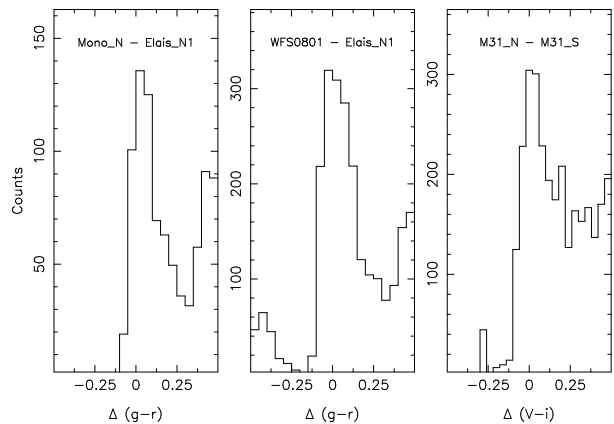}}
\caption[]{The three panels display  the distribution of counts in the
subtracted Hess diagrams shown in  the right-hand panels of Figures 3,
5 and  7.  Stars were selected in  the magnitude ranges $20.5  < g_0 <
22.0$ (for  the left-hand and middle  panels) and $20.5 <  V_0 < 22.0$
(for  the right-hand  panel), which  are  ranges in  which the  excess
population   in    the   subtracted   Hess-diagrams    is   relatively
uncontaminated.   The  abscissa displays  the  difference between  the
colour of  the stars and the  colour of the  ridge-line.  These narrow
peaks, centered  at a colour difference  close to zero,  show that the
excess  stellar population  closely follows  the ridge  line.  Fitting
these narrow  peaks with  a Gaussian function  (between ${\rm  -0.25 <
\Delta  Colour  < 0.25}$),  gives  a  colour  dispersion smaller  than
$\sigma=0.12$~magnitudes  for all  three cases.   Summing  under these
peaks, we find that the excess  population is detected with $S/N > 20$
in the  Mono-N field and  with $S/N >  30$ in the fields  WFS-0801 and
M31-N.}
\end{figure}

\section{Results and Conclusions}

The detection of a stellar population almost identical to that seen in
the SDSS  Monoceros fields shows  that this structure is  immense. The
M31  field is  $\sim 100^\circ$  away  in longitude  from the  S223+20
field,  towards the  diametrically opposite  side of  the  Galaxy. The
structure is also seen both below the Galactic plane (in the M31 field
and in  S200-24) and  above it (in  fields Mono-N,  WFS-0801, S183+22,
S218+22,  and  S223+20),  covering  a  vertical  range  of  more  than
$50^\circ$.   The fields  at  higher Galactic  latitude than  $|b|\sim
30^\circ$ did not  show up a similar CMD feature,  and neither was the
feature detected in the lower latitude ($b=-31^\circ$) M33 field.  The
structure appears to be confined close to the Galactic plane.

To  investigate the  difference in  Heliocentric distance  between the
fields,  we   compared  the  colour-magnitude   sequence  detected  by
\cite{newberg} in  the S223+20 field  to the three INT  fields Mono-N,
WFS-0801, and M31.   Due to the difference in  the photometric systems
between the surveys, conversions need to be made to shift measurements
in the  SDSS (g',r') system  to Vega-normalized (g,r) and  (V,i).  For
the (g,r) fields we used an South Galactic Pole region where there are
overlapping INT  and SDSS  data \cite[the latter  from the  Early Data
Release,][]{stoughton};    we   find    $(g-r)=0.21+0.86(g'-r')$   and
$g=g'+0.15-0.16(g-r)$.  The resulting transformed track of the S223+20
field requires an  offset of ${\rm -0.4~mag}$ to  match the Mono-N and
WFS-0801 fields (i.e.   stars in the Mono-N and  WFS-0801 features are
brighter).  To  convert to (V,i)  we took $V  = g - 0.03  - 0.42(g-r)$
\citep{windhorst}, and  used the WFS photometry in  the WFS-2240 field
to  find  the following  necessary  transformation:  $(g-i)  = 0.09  +
1.51(g-r)$. The offset of the S223+20 ridge-line was found to be ${\rm
-0.8~mag}$ for the northern M31 field (the feature in the southern M31
field has too  low S/N to fit).  Assuming that  the S223+20 feature is
$11\kpc$  distant \citep{newberg},  and  interpreting these  magnitude
offsets  as due  to distance  variations,  puts the  structure in  the
Mono-N, WFS-0801 and  M31 fields at a distance  of $\sim 9\kpc$, $\sim
9\kpc$   and    $\sim   8\kpc$   respectively.     The   corresponding
Galactocentric distances, in order of increasing Galactic longitude
\footnote{Please note that the photometric conversions, especially the
extrapolation  from (g',r')  on the  AB system  to (V,i)  on  the Vega
system,  may  have   significant  systematic  errors.   The  estimated
distances are therefore only  indicative.},
are:  $\sim 14\kpc$  in M31  (at $\ell=123^\circ$),  $\sim  16\kpc$ in
Mono-N   (at  $\ell=149^\circ$),   $\sim  16\kpc$   in   WFS-0801  (at
$\ell=180^\circ$), and $17.7\kpc$ in S223+20 (at $\ell=221^\circ$).

The substantial area of our M31 field allows a first estimation of the
scale-height of  this unexpected population.  To this  end we selected
stars in a banana-shaped region between  ${\rm 20.5 < V_0 < 21.5}$ and
${\rm  0.5+0.02(V_0-17.0)^2 <  (V-i)_0 <  0.8+0.02(V_0-17.0)^2}$.  The
density of these  sources is displayed as filled  circles in Figure~10,
and shows a rapid rise towards the Galactic plane. After subtracting a
``background'' level measured from the same colour-magnitude region in
the M33 field ($b=-31^\circ$), we performed a straight-line fit to the
logarithm (base-ten) of  the counts per square degree,  which yields a
slope of $0.081\pm0.004$ per  degree of Galactic latitude.  Taking the
distance  of $8\kpc$  derived above,  this implies  a scale  height of
$0.75\kpc\pm0.04\kpc$.  The  open circles  in Figure~10 show  a similar
selection, just  slightly bluer, which picks out  Galactic halo stars.
In contrast, this second  selection gives an almost flat distribution,
consistent  with   our  preconceptions  of  the  Halo   as  an  almost
spherically-distributed component.

\begin{figure}
\centerline{\psfig{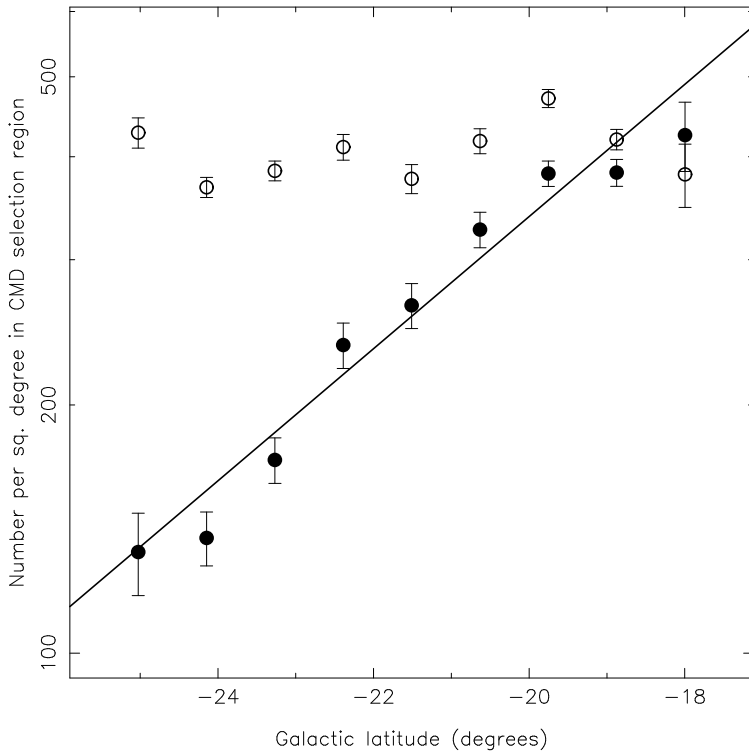}}
\caption[]{The filled circles show  the density of sources (plotted on
a logarithmic scale)  as a function of Galactic latitude  b in the M31
field that belong to the ``Monoceros-like'' main sequence feature that
is shown in Figures~6 and 8.  These sources have been selected between
${\rm  20.5 <  V_0  < 21.5}$,  and  with colours  lying between  ${\rm
0.5+0.02(V_0-17.0)^2   <   (V-i)_0   <  0.8+0.02(V_0-17.0)^2}$.    The
``background'' ($133.0\pm5.3$ counts per square degree, estimated from
the  M33  field)   has  been  subtracted  from  the   counts  of  this
``Monoceros-like'' population. A  straight-line fit through these data
is shown.  The  open circles show the corresponding  density of normal
Halo  sources, selected  with  ${\rm 21.0  <  V_0 <  22.0}$, and  with
colours     lying     between     ${\rm     0.5    <     (V-i)_0     <
0.4+0.02(V_0-17.0)^2}$. (No  background has been  subtracted from this
``Halo'' population).}
\end{figure}

The nature  of this structure remains  a puzzle.  It is  clear that it
cannot  be  related  to the  normal  thin  disk,  as it  lies  several
magnitudes below  the expected thin disk sequence.   The rapid decline
in the density of the feature  away from the Galactic plane also rules
out a  direct connection to the  halo.  This leaves the  thick disk as
the only normal  Galactic option. We used the  Galaxy starcounts model
of  \citet{ibata94} to  predict  the ${\rm  B-V}$, V  colour-magnitude
diagram of  the thick  disk in all  of the fields  investigated (${\rm
B-V}$ colours are close to ${\rm  g-r}$).  For the thick disk, we find
that the spread in V magnitude at constant colour has a FWHM exceeding
3 magnitudes in all of our  fields (we chose to sample 0.05 magnitudes
about ${\rm  B-V=1.0}$ which gives  the minimum spread  in magnitude).
The detected feature therefore cannot  be the standard thin disk, halo
or  even  thick  disk,  in agreement  with  \citet{newberg}.   Several
possibilities present themselves:

\begin{itemize}

\item[i.] One option is that  the various surveys probe small areas of
a gigantic ring  that encompasses the disk.  The ring  has a radius of
$15$--$20\kpc$,  a radial thickness  of $\sim  2\kpc$, and  a vertical
scale  height  of  $\sim  0.75\kpc$.   Due  to  the  presence  of  the
Magellanic Clouds and the Sagittarius dwarf galaxy, the ring is warped
and  non-circular, explaining the  variation in  Galactocentric radius
between some of the survey  fields.  The stellar mass of the structure
can be  estimated by taking  the surface density  in the M31  field as
representative of  the whole ring,  and assuming axisymmetry  and some
vertical  profile.  If  the  vertical profile  is  exponential, as  is
suggested by  the data displayed in  Figure~10, we find  that the total
stellar mass of the structure is $M=10^9\msun$, whereas if the surface
density remains constant below  $|b|=18^\circ$, the mass is $M=2\times
10^8 \msun$ (for these estimates we have taken the luminosity function
of   \citealt{jahreiss}   and    the   mass-luminosity   relation   of
\citealt{henry}).  A  possible mechanism  to form such  a ring  may be
repeated warpings  of the outer  disk.  In the Andromeda  Galaxy (seen
almost   edge-on),  the   outer   stellar  disk   has  a   complicated
non-axisymmetric   shape  \citep{ferguson02},   with   large  vertical
deviations suggestive of ancient warps  which now no longer follow the
warp seen in the gas.  The possible Ring seen in the outer disk of the
Milky Way may be of  similar origin.  One would expect such structures
to eventually mix entirely, leaving a flared outer disk, rather than a
radially thin ring.  Either there  is another factor at play, such as,
for instance,  a strong interaction with the  Sagittarius dwarf galaxy
\citep{ibata98b}  or the Magellanic  Clouds \citep[e.g.,][]{tsuchiya},
or the  ring is  a recent phenomenon  (note however, that  the stellar
constituents could be older than  the structure).  If the structure is
a perturbation, the wave may  have been amplified to the current large
vertical extent (in the manner  of a whip) in passing outwards through
the disk to regions of progressively lower density.

The    low-latitude    HVC    field    \citep{lewis},    located    at
($\ell=327^\circ$,  $b=-15^\circ$), is not  in contradiction  with the
interpretation of the structure as a Galactic ring.  Assuming that the
ring has a radius of $18\kpc$,  implies a distance of $24\kpc$ in this
direction, more  than twice  the distance to  the fields in  which the
``Monoceros''-like CMD structure has been detected.  The HVC field may
simply lie  too far  below the Galactic  plane (in that  direction) to
detect the Ring (the field  also covers a relatively small solid angle
of sky).

The reason the Ring may not  have been discovered before is due to its
low surface  density. Very large areas  need to be  surveyed to detect
it.  It would  also be  difficult to  detect with,  for  instance, the
poorer photometry available from photographic plates.

\item[ii.]  Another  possibility is  that the surveys  probe different
regions of the disrupted tidal stream of an accreted satellite galaxy.
Halo satellites  are expected to  have high eccentricity  orbits, with
apocentre  to  pericentre  ratio   of  $\sim  4$  \citep{bosch}.   For
sufficiently  massive satellites  ($\simgt 10^9\msun$),  the continual
braking effect of  dynamical friction with the Galactic  halo and disk
can eventually lead  to the decay of the  orbit.  However, simulations
show that  satellite orbits are not readily  circularized by dynamical
friction  \citep{bosch,colpi},  so it  is  hard  to  explain the  near
circular orbit required by this  scenario. A second problem relates to
the small  distance spread along the  line of sight.  As  the orbit of
the  satellite  decays,  constituent  stars  will  be  lost  by  tidal
disruption. Over time, the  tidally removed stars become phased-mixed,
spreading out  over the  orbit, and occupying  the region  between the
orbit pericentre and apocentre.   The fact that the ``Monoceros''-like
population is  seen confined in  a narrow distance interval  along the
line  of  sight therefore  also  argues  against  this scenario.   The
problem may  be alleviated if the progenitor  satellite disrupted only
recently.   A  further  concern  with  this scenario  comes  from  the
disruptive  effect  of  the  satellite  on  the  Milky  Way.   Current
cosmological simulations suggest that galaxy satellites have their own
massive dark matter mini-haloes  \citep{stoehr}, a possibility that is
supported  by  the high  mass  to  light  ratios inferred  from  small
Galactic  dwarf  spheroidals  \citep{mateo},  and  from  the  survival
requirement   of  the   Sgr  dwarf   galaxy  \citep{ibata97,ibata98a}.
Including the dark matter halo of  a satellite that has a stellar mass
in   the   range   $2\times10^8   \msun$  to   $10^9\msun$   increases
significantly the potential  damage to the disk of  the Milky Way, and
raises the  question of  whether the thin  disk would survive  such an
encounter \citep[see, e.g.,][]{toth,velazquez}.

\item[iii.]   Another possibility  is that  we are  seeing part  of an
outer  spiral  arm,  or  various arm  fragments.   \citet{davies}  has
identified a variety  of spiral features in the  outer Galaxy via 21cm
emission, extending out to a radius  of $\sim 25 \kpc$.  Many of these
structures lie in directions where  we have also detected an anomalous
stellar  component,  leading  one  to  speculate  if  these  could  be
associated stellar  arms.  Indeed, if  an underlying global  mode were
responsible  for  driving  the  spiral  structure,  it  would  not  be
unexpected to  find an older  stellar population tracing out  the same
pattern   as  the   gas   \citep[e.g.][]{rix,thornley}.   The   narrow
line-of-sight  thickness of the  structures stands  in favour  of this
hypothesis. The thickness of  the stellar component may be problematic
for the spiral  arm hypothesis however, although one  might be able to
appeal to warping to bring the bulk of the stars out of the plane.

\item[iv.]  The  beating motion  of an asymmetric  Galactic component
(such as the bar) induces  resonances in the disk component.  However,
these resonances occur very close to  the plane of the disk, and stars
with significant vertical motions are  unlikely to partake in a strong
resonance. It  seems implausible, therefore,  that a resonance  is the
cause of the detected structure.

\end{itemize}

The  photometric  information  presented  here is  not  sufficient  to
discriminate between  the first three  scenarios.  Further photometry,
especially at low Galactic latitude will be invaluable if we are to be
able to properly  follow the structure, and ascertain  whether it is a
Galactic  Ring,  an  inhomogeneous  mess  due  to  ancient  warps  and
disturbances,  or  part of  a  disrupted  satellite  stream.  If  this
manifestly old population  turns out to be the  outer stellar disk, it
will pose a very interesting challenge to galaxy formation models that
predict inside-out assembly. Alternatively,  if it transpires that the
structure  is  due to  a  disrupted  satellite  whose orbit  has  been
circularised and accreted along with its cargo of dark matter onto the
Disk, it  will provide a  unique first-hand opportunity  to understand
the effect of massive accretions on to the inner regions of galaxies.

\newcommand{\mnras}{MNRAS}
\newcommand{\nat}{Nature}
\newcommand{\araa}{ARAA}
\newcommand{\aj}{AJ}
\newcommand{\apj}{ApJ}
\newcommand{\apjl}{ApJ}
\newcommand{\apjs}{ApJSupp}
\newcommand{\aap}{A\&A}
\newcommand{\aaps}{A\&ASupp}
\newcommand{\pasp}{PASP}

\onecolumn

\begin{table*}
\centering
\begin{minipage}{140mm}
\caption{Summary   of  appropriate   WFC   observations,  ordered   in
descending  $|b|$.    The  final  column  lists   whether  the  excess
``Monoceros'' population is detected in the field. The significance of
the detection is listed for those fields where it has been possible to
calculate this  parameter. (The significance cannot  be calculated for
all of the INT fields due  to a lack of suitable background fields, or
of a sufficiently accurate Galaxy  model). We refrain from listing the
stellar density in these fields, as the values cannot be compared in a
simple manner,  due to the  different photometric systems used  in the
various surveys  (Figure~10 provides our current best  estimate of the
vertical structure of the population).}
\begin{tabular}{lccccccc} \hline \hline
Field            & RA (J2000)   & Dec (J2000) & $\ell$    &      b    &Area (sq. deg.)& Bands &  Detection?   \\ \hline
Equatorial survey&  22 to 3     &   0:00      & 60 to 180 & -40 to -60& $\sim 20$     & g,r   &   NO          \\
WFS-2240         &    22:40     &  +0:00      &    69     &    -48    & 9.0           & g,r   &   NO          \\
Elais-N1         &    16:13     & +55:16      &    85     &    +44    & 9.0           & g,r   &   NO          \\
Elais-N2         &    16:36     & +41:01      &    64     &    +41    & 9.0           & g,r   &   NO          \\
M33              &     1:34     & +30:40      &   134     &    -31    & 4.8           & V,i   &   NO          \\
WFS-0801         &    08:02     & +40:19      &   180     &    +30    & 7.0           & g,r   &  $S/N > 30$   \\
M31 comparison 2 &    23:50     & +35         &   109     &    -26    & 1.2           & V,i   &   NO          \\
S200-24 
\footnote{SDSS field, Newberg et al. (2002)}
                 &     5:00     &  +0:00      &   199     &    -25    & $\sim 5$      & g',r' &  YES          \\
M31-S            &     0:43     & +38:45      &   122     &    -24    & 8.25          & V,i   &  YES          \\
S183+22  $^{a}$  &     7:22     & +35         &   183     &    +22    & $\sim 5$      & g',r' &  YES          \\
S218+22  $^{a}$  &     8:16     &  +6         &   218     &    +22    & $\sim 5$      & g',r' &  YES          \\
M31 comparison 1 &    22:27     & +31         &    90     &    -21    & 1.2           & V,i   &  PERHAPS      \\
M31-N            &     0:47     & +43:22      &   123     &    -19    & 6.5           & V,i   &  $S/N > 30$   \\
Mono-N           &     6:03     & +64:43      &   149     &    +20    & 1.2           & g,r   &  $S/N > 20$   \\
S223+20 $^{a}$   &     8:00     &  +0:00      &   221     &    +15    & $\sim10$      & g',r' &  YES          \\
HVC
\footnote{Based on AAT WFI data, Lewis et al. (2002)}
                 &    17:13     & -64:38      &   327     &    -15    & 0.25          & g,r   &   NO          \\
\hline    
\end{tabular}
\end{minipage}
\end{table*}

\end{document}